\documentstyle[eqsecnum,aps,epsfig]{revtex}

\begin{document}
\draft


\title{Comment on ``Long-range electrostatic interactions between like-charged colloids: Steric and 
confinement effects'' 
}
\author{Eduard M. Mateescu 
}
\address{University of California Santa Barbara, Santa Barbara, California 93106
}
\date{6 October 2000}
\maketitle

\begin{abstract}
In a recent study [Phys.\ Rev.\ E\ {\bf 60}, 6530 (1999)], Trizac and Raimbault showed that the effective pair interaction between like charged colloids immersed in a cylindrically confined electrolyte remains repulsive even when the size of the micro-ions or the finite longitudinal extension of the confining cylinder are taken into account. Contrary to their claim, we argue that the case of finite longitudinal confinement doesn't always generate repulsive interactions and to illustrate this point we also provide a simple example. 
\end{abstract}



There have been a number of recent theoretical efforts towards understanding the mechanism behind the long range attraction that is experimentally observed \cite{frad,grier2,carb,lars,grier3} when like charged colloids are immersed in a confined electrolyte. The fact that this interaction returns to a purely repulsive one as soon as the confining surface is removed \cite{lars,grier1}, suggested that the attractive part could be obtained within the framework of the Poisson-Boltzmann theory by properly taking into account the influence of the confining surface. However, Neu \cite{neu} and Sader and Chan \cite{chan}, using a general model of cylindrical confinement, have rigorously proven that this is not the case.

In their paper, Trizac and Raimbault \cite{triz} extended the aforementioned studies \cite{neu,chan} to include the influence brought upon the sign of the interaction by the size of the micro-ions and the finite longitudinal extension of the confining cylinder (case hereafter referred to as complete confinement). Although we agree with their conclusion that the interaction remains repulsive when the specific size of the micro-ions is taken into account through a modified Poisson-Boltzmann equation, we don't agree with a similar conclusion  for the case of complete confinement. In this comment, we show that this controversy is brought about by an inaccuracy in their calculation and that in general, colloids immersed in a completely confined electrolyte don't always repel. We also provide a simple example to illustrate this point.

To be more specific, the issue has to do with the use of Green's theorem in two dimensions ($2D$). If $W$ is a closed domain in ${\bf R^2}$ (for example in the $xy$ plane) and $\phi(x,y)$ and $\psi(x,y)$ are scalar fields defined on ${\bf R^2}$, then Green's theorem in $2D$ is:

\begin{equation}
\int_{W} \!dxdy\,\left[\left(\nabla_{xy}\phi\right)\!\cdot\!\left(\nabla_{xy}\psi\right)+\phi\,\nabla_{xy}^2\psi\right]=\oint_{\partial W}\!d\ell\ \phi\, \left({\bf \hat{n}}\!\cdot\!\nabla_{xy}\psi\right),
\label{green}
\end{equation}

where ${\bf \hat{n}}$ (which is contained in the $xy$ plane) represents the outer unit normal to the boundary $\partial W$ $\!$ of $W$ and $\nabla_{xy}$ is the $2D$ gradient operator.

With this in mind and making use of the same notation and assumptions as in \cite{triz}, we obtain a different expression for Eqs.~(14) and (15) in that paper, where Green's theorem was inaccurately applied. For clarity, let us first separate the $z$ component of the gradient in the equation that generates the results $(14)$ and $(15)$ in \cite{triz}:

\begin{eqnarray} 
&&\int_{Oxy}\!dxdy\ {\bf E}_{z=L}\!\cdot\!\left({\bf D}_{z=0}-{\bf D}_{z=L}\right)
\label{prost}\\
&&=\int_{Oxy}\!dxdy\, \left(-\nabla\psi\right)_{z=L}\!\cdot\!\left[ \epsilon\left(-\nabla\psi\right)_{z=0} - \epsilon\left(-\nabla\psi\right)_{z=L} \right]
\label{pic}\\
&&=\int_{Oxy}\!dxdy\left\{\epsilon\left(-\nabla_{xy}\psi\right)_{z=L}\!\cdot\!\left[-\left(\nabla_{xy}\psi\right)_{z=0} + \left(\nabla_{xy}\psi\right)_{z=L}\right] - \epsilon\left(\frac{\partial\psi}{\partial z}\right)_{z=L}^2 \right\},
\label{second}
\end{eqnarray}

where $\nabla$ is the $3D$ gradient operator and $\psi$ possesses mirror symmetry with respect to the plane $z=0$, as in \cite{triz}. The reader should remember that the notation $\int_{Oxy}$ actually stands for the two integrals over the domains of the $xy$ plane that are interior and exterior, respectively, to the confining cylinder. The dimensionality of these integrals requires the use of the $2D$ version of Green's theorem, unlike it was inaccurately done in \cite{triz}, where the $3D$ version was employed instead. This issue is clarified if we consider the more explicit representation shown in Eq.~(\ref{second}), where we can correctly apply the $2D$ version of Green's theorem [as stated in Eq.~(\ref{green})] to its first term to get:

\begin{eqnarray} 
&&\int_{\partial\Sigma}\!d\ell \left(D_n-D'_n\right)_{z=L} \left(-\psi_{z=o}+\psi_{z=L}\right) - \int_{Oxy}\!dxdy\left[\epsilon\left(\nabla_{xy}^2\psi\right)_{z=L} \left(\psi_{z=o}-\psi_{z=L}\right)+\epsilon\left(\frac{\partial\psi}{\partial z}\right)_{z=L}^2 \right]
\label{a} \\
&&=\sigma\!\int_{\Sigma}\!dS\,E_z-\int_{Oxy}\!dxdy\left[\left(\psi_{z=o}-\psi_{z=L}\right)\frac{\partial P}{\partial\psi}\left(\psi_{z=L}\right) - \left(\psi_{z=o}-\psi_{z=L}\right)\epsilon\left(\frac{\partial^2\psi}{\partial z^2}\right)_{z=L}\!+\epsilon\left(E_{z}\right)_{z=L}^2 \right],
\label{b}
\end{eqnarray}

where in going from Eq.~(\ref{a}) to Eq.~(\ref{b}) we used the dielectric boundary condition $D'_n-D_n=\sigma$ (and the fact that $\int_{\partial\Sigma}\!d\ell\left(\psi_{z=0}-\psi_{z=L}\right)=\int_{\Sigma}\!dS\,E_{z} $) and also formed the $3D$ Laplacian and employed Eq.~(10) in \cite{triz}.

In fact, Eqs.~(\ref{a}) and (\ref{b}) are the correct forms for Eqs.~(14) and (15), respectively, in \cite{triz}. After adding all the terms, the correct expression for the $z$ component of the force (Eq.~(16) in \cite{triz}) should read:

\begin{eqnarray}
F_z=&&\int_{Oxy}\!dxdy \left[ P\left(\psi_{z=0}\right)-P\left(\psi_{z=L}\right)-\left(\psi_{z=0}-\psi_{z=L}\right)\frac{\partial P}{\partial\psi}\left(\psi_{z=L}\right) \right] \nonumber\\
&&+ \int_{Oxy}\!dxdy\ \frac{\epsilon}{2} \left({\bf E}_{z=0}-{\bf E}_{z=L}\right)^2 + \int_{Oxy}\!dxdy\left(\psi_{z=0}-\psi_{z=L}\right)\epsilon\left(\frac{\partial^2\psi}{\partial z^2}\right)_{z=L}.
\label{c}
\end{eqnarray}

Although one can use the same argument as in \cite{triz} to prove that the first two integrals in  Eq.~(\ref{c}) are always positive, the third one (which is different from that obtained by Trizac and Raimbault) can in fact be negative and sometimes overcome the first two. Before we show how this can happen, let us mention another equivalent expression for $F_{z}$, which can be obtained by recasting in a different manner the second integrand of Eq.~(13) in \cite{triz}, namely $\left({\bf D}\!\cdot\!{\bf E}\right)_{z=0}-\left({\bf D}\!\cdot\!{\bf E}\right)_{z=L}=-\epsilon\left({\bf E}_{z=0}-{\bf E}_{z=L}\right)^2-2\,{\bf E}_{z=0}\!\cdot\!\left({\bf D}_{z=L}-{\bf D}_{z=0}\right)$ and then following the same steps used in deriving Eq.~(\ref{c}):

\begin{eqnarray}
F_z=&&-\int_{Oxy}\!dxdy \left[P\left(\psi_{z=L}\right)-P\left(\psi_{z=0}\right)-\left(\psi_{z=L}-\psi_{z=0}\right)\frac{\partial P}{\partial\psi}\left(\psi_{z=0}\right) \right] - \int_{Oxy}\!dxdy\ \frac{\epsilon}{2} \left({\bf E}_{z=0}-{\bf E}_{z=L}\right)^2 \nonumber\\
&& + \int_{Oxy}\!dxdy\ \epsilon \left(E_z\right)_{z=L}^2 - \int_{Oxy}\!dxdy\left(\psi_{z=L}-\psi_{z=0}\right)\epsilon\left(\frac{\partial^2\psi}{\partial z^2}\right)_{z=0}.
\label{d}
\end{eqnarray}

The reader can easily check that Eqs.~(\ref{c}) and (\ref{d}) are equivalent, by subtracting them from one another and then using Eq.~(10) in \cite{triz}, Green's theorem in $2D$ and the dielectric boundary condition to prove that the result is zero.  

\begin{figure}[hbtp]

\centerline{\epsfig{file=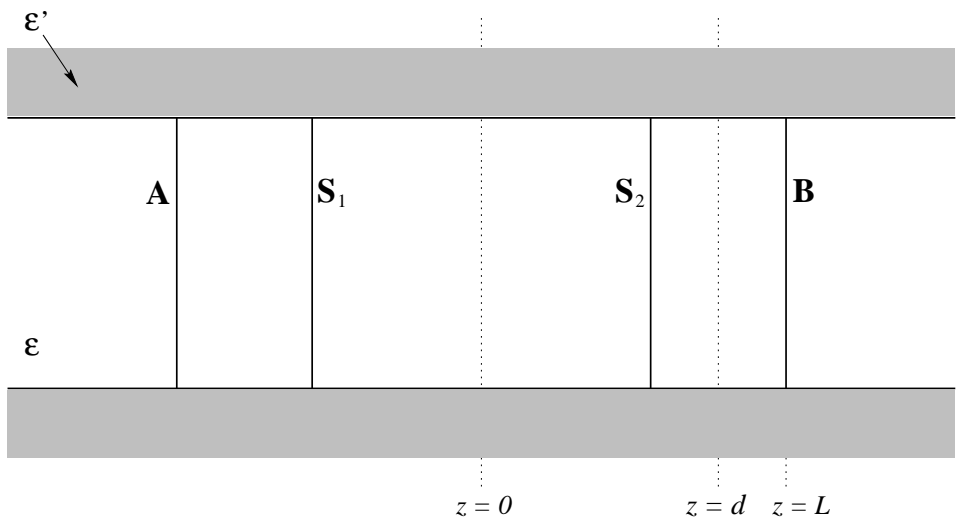,height=5in,angle=270}}
\vskip .5cm

\caption{Longitudinal section along the confining cylinder showing the two particles ${\bf S}_1$ and ${\bf S}_2$ and the walls {\bf A} and {\bf B} responsible for the complete confinement. The plane $z=d$ is midway between the particle ${\bf S}_2$ and the wall {\bf B}.}
\label{fig}
\end{figure}

We can now propose a simple example in order to prove that the complete confinement doesn't always generate repulsive interactions. As before, we assume that the particles are immersed in an electrolyte of permittivity $\epsilon$, confined by an infinite and uniformly charged cylindrical surface (of arbitrary cross section), while the medium outside the cylinder has a permittivity $\epsilon'$ and also contains an electrolyte (most general case). The only requirement is that $\psi$ possesses mirror symmetry with respect to the plane $z=0$. In our example the particles ${\bf S}_1$ and ${\bf S}_2$ are taken to be flat surfaces with the same surface charge density $\sigma$ as the confining cylinder while the micro-ions are able to equilibrate at all times (by being in contact with a reservoir and possibly permeating through the colloidal particles). The walls {\bf A} and {\bf B} that realize the complete confinement are surrounded on both sides by electrolyte and also have the same uniform surface charge density $\sigma$ as the confining cylinder (this way the complete confinement is realized by a surface of constant $\sigma$). To compute the z component of the force on the colloid ${\bf S}_2$, we follow the same procedure outlined in \cite{triz} only that instead of choosing the surface $\Sigma_{L}$ to integrate over in Eq.~(11) that defines $F_{z}$ in \cite{triz}, we choose the surface $\Sigma_{d}$, midway between the colloid ${\bf S}_2$ and the wall {\bf B} (see Fig.~\ref{fig}). Keeping in mind the correction made in this comment, the result ends up to be identical to Eq.~(\ref{d}), with the only exception that all the indices $z=L$ are replaced by $z=d$:

\begin{eqnarray}
F_{z}=&&-\int_{Oxy}\!dxdy \left[P\left(\psi_{z=d}\right)-P\left(\psi_{z=0}\right)-\left(\psi_{z=d}-\psi_{z=0}\right)\frac{\partial P}{\partial\psi}\left(\psi_{z=0}\right) \right] - \int_{Oxy}\!dxdy\ \frac{\epsilon}{2} \left({\bf E}_{z=0}-{\bf E}_{z=d}\right)^2 \nonumber\\
&&+ \int_{Oxy}\!dxdy\ \epsilon \left(E_{z}\right)_{z=d}^2 - \int_{Oxy}\!dxdy\left(\psi_{z=d}-\psi_{z=0}\right)\epsilon\left(\frac{\partial^2\psi}{\partial z^2}\right)_{z=0}.
\label{f2z}
\end{eqnarray} 

It is easy to see that the last two integrals in Eq.~(\ref{f2z}) approach zero as the separation $2L$ between the walls is increased to infinity and the distance $2(L-d)$ between each particle and the adjacent wall is kept constant. Indeed, in this situation, $\psi$ becomes symmetric with respect to the plane $z=d$ (due to our choice for the shape of the colloidal particle) and therefore $\left(E_z\right)_{z=d}\longrightarrow 0$, while at $z=0$ (infinitely far away from the particles and the walls), the problem of solving for $\psi$ becomes essentially a $2D$ one (independent of $z$) and therefore $\partial^n\psi/\partial z^n \longrightarrow 0$ for any $n \geq 1$. At the same time, as argued in \cite{triz}, the first two integrals in Eq.~(\ref{f2z}) are always positive (the first integrand is positive due to the global convexity of $P(\psi)$) and moreover, reach some finite asymptotic values in this limit case. Since both integrals contribute with a minus sign in Eq~(\ref{f2z}), it becomes clear that $F_{z}<0$ when $L\longrightarrow\infty$ and $L-d$ is constant.  

Therefore, within the approximations of our example, we conclude that for any fixed distance $2(L-d)$ between the colloidal particle and the adjacent wall, there is a finite, maximum separation between the walls $2L_{0}$, at which the particles are in equilibrium ($F_z=0$) and beyond which $F_{z}$ becomes negative and remains so as $L$ increases. Although this is in contradiction with the result obtained earlier by Trizac and Raimbault \cite{triz} that $F_{z}$ should always be positive under complete confinement, we have shown that the controversy stems from an inaccuracy in their calculation. It is therefore possible that completely confined colloidal particles don't always repel, mainly due to the stronger interaction with the walls (or conceivably other particles \cite{goulding}) that are responsible for the finite longitudinal confinement and break the translational symmetry of the confined space. Although this type of calculation is not suitable to yield an estimate on $L_{0}$, it would nevertheless be of high interest if this could be done through computer simulations.

\acknowledgments

It is a pleasure to acknowledge stimulating discussions with M. N. Tamashiro, C. Jeppesen and P. Pincus. This work was supported by the DMR Program of the National Science Foundation under Awards DMR99-72246 and DMR96-32716.


\end{document}